\documentclass[review,number,sort&compress]{elsarticle} 
\usepackage{lineno}
\usepackage{graphicx}  
\usepackage{gensymb}
\usepackage{float}
\usepackage{booktabs}
\usepackage{natbib}
\usepackage{xcolor}
\usepackage[english]{babel}

\begin{document}
\title{Analysis of the background on cross section measurements with the MAGNEX spectrometer: \\the ($^{20}$Ne,$^{20}$O) Double Charge Exchange case}

\author{S.~Calabrese$^{a,b}$ \corref{cor}}
\cortext[cor]{Corresponding author. Address: LNS/INFN, Via Santa Sofia 62, 95123, Catania, Italy. Phone Office: +39 095542809. e-mail: calabrese@lns.infn.it \\ \copyright  2020. This manuscript version is made available under the CC-BY-NC-ND 4.0 license https://creativecommons.org/licenses/by-nc-nd/4.0/}
\author{F.~Cappuzzello$^{a,b}$, D.~Carbone$^{a}$, M.~Cavallaro$^{a}$, C.~Agodi$^{a}$, D.~Torresi$^{a}$, L.~Acosta$^{c}$, D.~Bonanno$^{d}$, D.~Bongiovanni$^{a}$, T.~Borello-Lewin$^{e}$, I.~Boztosun$^{f}$, G.~A.~Brischetto$^{a,b}$, D.~Calvo$^{g}$, I.~Ciraldo$^{a,b}$, N.~Deshmukh$^{a}$, P.N.~de~Faria$^{h}$, P.~Finocchiaro$^{a}$, A.~Foti$^{b,d}$, G.~Gallo$^{b,d}$, A.~Hacisalihoglu$^{a,i}$, F.~Iazzi$^{g,j}$, R.~Introzzi$^{g}$, L.~La Fauci$^{a,b}$, G.~Lanzalone$^{a,k}$, R.~Linares$^{h}$, F.~Longhitano$^{d}$, D.~Lo Presti$^{b,d}$, N.~Medina$^{e}$, A.~Muoio$^{a}$, J.R.B.~Oliveira$^{e}$, A.~Pakou$^{l}$, L.~Pandola$^{a}$, F.~Pinna$^{g,l}$, S.~Reito$^{d}$, G.~Russo$^{b,d}$, G.~Santagati$^{a}$, O.~Sgouros$^{a,l}$, S.O.~Solakci$^{f}$, V.~Soukeras$^{a,l}$, G.~Souliotis$^{m}$, A.~Spatafora$^{a,b}$, S.~Tudisco$^{a}$, V.A.B.~Zagatto$^{h}$\newline \\for the NUMEN collaboration\\} 
 
\address{$^a$ INFN, Laboratori Nazionali del Sud, Catania, Italy\\ $^b$ Dipartimento di Fisica e Astronomia "E.Majorana", Universit\`a di Catania, Catania, Italy\\ $^c$ Instituto de F\' isica, Universidad Nacional Aut\' onoma de M\' exico, Mexico City, Mexico \\ $^d$ INFN, Sezione di Catania, Catania, Italy\\
$^e$ Departamento de F\' isica Nuclear, Instituto de F\' isica, Universidade de S\~ao Paulo, S\~ao Paulo, Brazil \\ $^f$ Department of Physics, Akdeniz \"Universitesi, Antalya, Turkey\\ $^g$ INFN, Sezione di Torino, Turin, Italy\\$^h$ Instituto de F\' isica, Universidade Federal Fluminense, Niter\' oi, Rio de Janeiro, Brazil\\ $^i$ Institute of Natural Sciences, Karadeniz Teknik \"Universitesi, Trabzon, Turkey \\ $^j$  Dipartimento Scienza Applicata e Tecnologia, Politecnico di Torino, Turin, Italy \\ $^k$ Facolt\`a di Ingegneria e Architettura, Universit\`a di Enna "Kore", Enna, Italy \\$^l$ Department of Physics and HINP, University of Ioannina, Ioannina, Greece \\ $^m$ Department of Chemistry, National and Kapodistrian University of Athens and HINP, Athens, Greece}

\begin{abstract}
The MAGNEX magnetic spectrometer is used in the experimental measurements of Double Charge Exchange and Multi-Nucleon Transfer reactions induced by heavy ions within the NUMEN project. These processes are characterized by small cross sections under a large background due to other reaction channels. Therefore an accurate control of the signal to background ratio is mandatory. In this article, the determination of the MAGNEX spectrometer background contribution on cross section measurements is presented by applying a suitable analysis to quantify the limits of the adopted particle identification technique. The method is discussed considering the $^{116}$Cd($^{20}$Ne,$^{20}$O)$^{116}$Sn Double Charge Exchange reaction data, however it can be applied to any other reaction channel of interest.     
\end{abstract}
\date{}

\maketitle 
  
\section{Introduction}
The MAGNEX large acceptance magnetic spectrometer, installed at Laboratori Nazionali del Sud (INFN-LNS) in Catania, is a powerful and versatile device for the study of nuclear reactions of several systems. It has been used in a variety of nuclear physics researches at bombarding energies between the Coulomb barrier and the Fermi energy \cite{Oliveira_2013, cappuzzello2015, PhysRevC.93.064323, PhysRevC.95.034603, PhysRevC.95.054614, 2015NatCo, PhysRevC.100.034620, PhysRevC.97.034616, PhysRevC.97.054608}. An accurate description of the facility and its operation procedure together with the discussion of some relevant recently obtained results is available in Ref. \cite{magnex_review}. Currently, its use is strongly connected to the NUMEN project, whose experimental goals rely on the accurate absolute cross sections measurements of heavy-ion induced double charge exchange (DCE) reactions in view of their connection to neutrinoless double beta decay physics as well as the competitive multi-nucleon transfer and quasi-elastic processes. The details of the project are beyond the aims of this work and are widely discussed in Ref. \cite{numen2018}. A general issue of these studies is represented by the very low cross sections observed for DCE reactions, typically ranging from few nb to tens $\mu$b, depending on the particular case \cite{cappuzzello2015, naulin, blomgren_rcnp, takakia}. In principle, these cross section values could be comparable to the limits of typical experimental set-up, so a background study is needed. Moreover, the analysis of such quantity is essential to address the suitable detection technology in the view of the next upgrade of the INFN-LNS facility \cite{Iazzi,Calabretta,Tudisco,Muoio,Zero_submitted} driven by NUMEN.
\newline
Here we present a general method to estimate the unavoidable particle identification contaminations affecting the absolute cross section measurement. It has been developed analyzing the $^{116}$Cd($^{20}$Ne,$^{20}$O)$^{116}$Sn reaction data at 15.3 AMeV for which we have already shown promising results \cite{calabrese2018} but it can be directly applied to any other projectile-target system and reaction channel explored with the same experimental set-up.

\section{Particle identification}
\begin{figure*}[]
\centering
\includegraphics[width=0.50\textwidth]{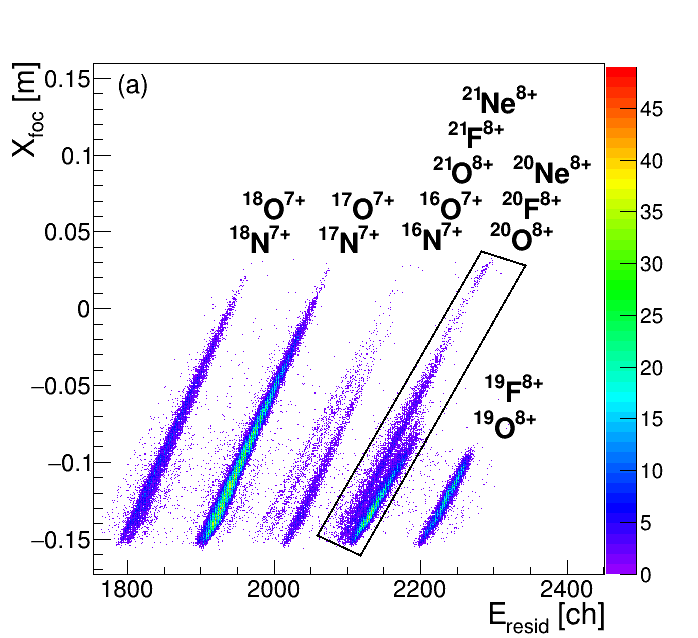} \\
\includegraphics[width=0.50\textwidth]{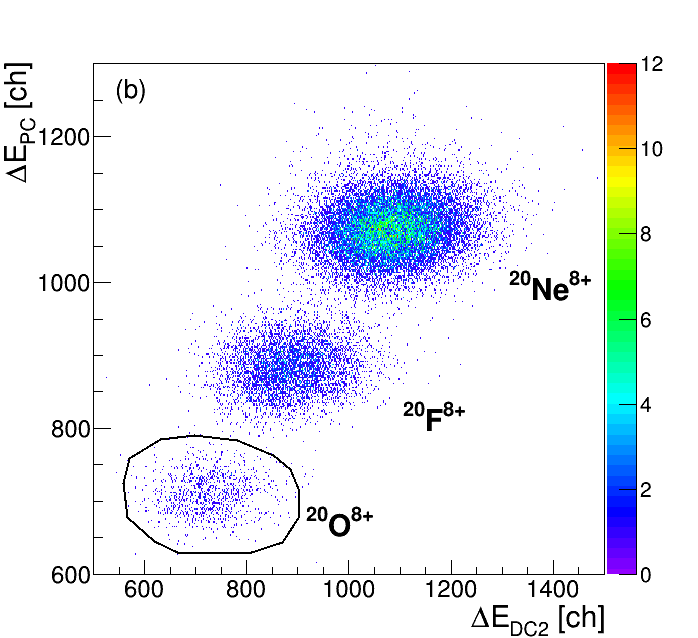} \\
\includegraphics[width=0.50\textwidth]{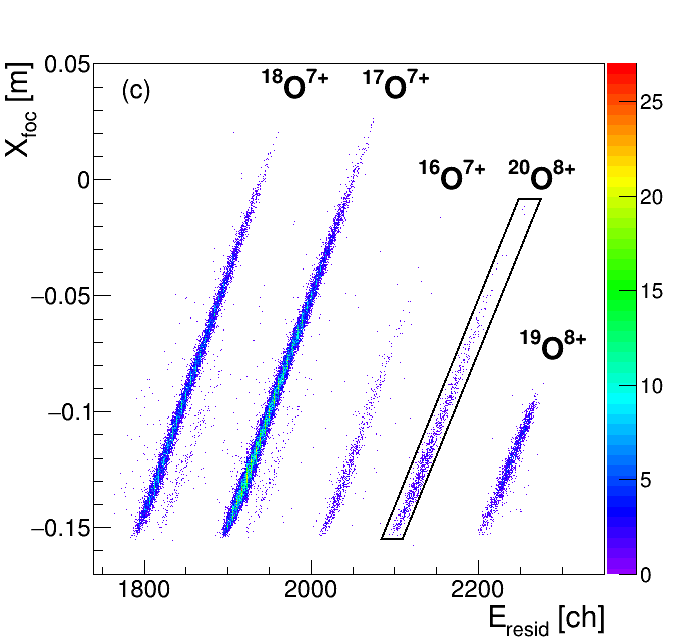}
\centering \caption{(Colour online) Example of the methodology used for the particle identification of $^{20}$O$^{8+}$ ions in one silicon detector (No. 29). Panel (a): Typical \textit{$X_{foc}$ vs }$E_{resid}$ representation for all the ungated events. A first raw selection in mass number and charge state for the $^{20}$F$^{8+}$, $^{20}$O$^{8+}$ and $^{20}$Ne$^{8+}$ ions is indicated by the graphical contour reported. Panel (b): \textit{$\Delta E_{PC}$ vs $\Delta E_{DC2}$} plot for the events gated by the graphical selection shown in panel (a). The identification in atomic number of $^{20}$O$^{8+}$ is shown by the graphical contour. Panel (c): \textit{$X_{foc}$ vs }$E_{resid}$ representation for the events gated only by the graphical cut shown in panel (b). The fine selection in mass number and charge state for the $^{20}$O$^{8+}$ ions is shown by the graphical contour.}  
\label{Fig:1}
\end{figure*}
The MAGNEX focal plane detector (FPD) is a three dimensional gas-filled tracker detector, divided into multiple proportional regions where the energy losses ($\Delta E_{DC1,DC2,DC3,DC4,PC}$) as well as the horizontal and vertical  positions (\textit{$X_{foc}$}, \textit{$Y_{foc}$}) and angles (\textit{$\theta_{foc}$}, \textit{$\phi_{foc}$}) of the tracks are measured. The FPD is completed by fifty-seven silicon pad detectors to measure the ion residual energies (\textit{$E_{resid}$}). The different silicon pad detectors, arranged in nineteen columns of three each, are placed along the dispersive - horizontal - direction of the focal plane, thus intercepting ejectiles of different kinetic energies according to their magnetic rigidities. In particular, due to the action of the MAGNEX quadrupole field, the central silicon detectors of the different columns intercept more ejectiles than the upper and lower ones of the same column. For such reason in the following analysis only central silicon pad detectors will be considered. 
The typical particle identification (PID) technique adopted with the MAGNEX spectrometer is described in details in Ref.~\cite{cappuzzello2010}. The atomic number (\textit{Z}), the mass number (\textit{A}) and the charge state (\textit{q}) are determined for each ion crossing the FPD \cite{cavallaro2012, Torresi_inpreparation}.  
In the present work, as shown in Fig.~\ref{Fig:1}(b) for the $^{20}$O$^{8+}$ identification, \textit{Z} is selected exploiting the correlation between the energy losses at two different sections of the FPD ionization chamber (\textit{$\Delta E_{PC}$, $\Delta E_{DC2}$}) corrected by the different path lengths in the gas, consequence of different incident angles at the FPD. \textit{A} and \textit{q}, instead, are typically deduced performing a precise reconstruction of the ions kinetic energy, as demonstrated in Ref.~\cite{cappuzzello2010}. However, in the experimental conditions of the NUMEN reactions as the one here described, which involve oxygen, fluorine and neon ions, a high mass resolution is not necessary and the identification procedure is successfully performed using the correlation between the measured position at the focal plane in the ejectiles dispersive direction (\textit{$X_{foc}$}) and their residual energy (\textit{$E_{resid}$}) measured by the stopping silicon detector after crossing the FPD gas section. As discussed in Ref.~\cite{numen2018, CAVALLARO2020334}, such two quantities are proportional through a factor $\frac{\sqrt{m}}{q}$ depending on the ejectile mass \textit{m} and charge state \textit{q}, see Figs.~\ref{Fig:1}(a), 1(c). \newline Thanks to the achieved resolution - $\frac{1}{160}$ in \textit{A} and $\frac{1}{48}$ in \textit{Z} \cite{cappuzzello2010} - the amount of misidentified ions for the direct reaction channels explored so far with the MAGNEX facility - characterized by mb down to $\mu$b cross sections - was negligible. However, for the purposes of NUMEN, such investigation becomes necessary due to the low cross sections expected for the channels of interest which are typically as low as a few nb. 
\newline
In the experiment a beam of $^{20}$Ne$^{10+}$ ions, extracted by the K800 Superconducting Cyclotron, impinged at 306 MeV incident energy on a 1360$\pm$70 $\mu$g/cm$^{2}$ $^{116}$Cd target coupled to a 990$\pm$50 $\mu$g/cm$^{2}$ natural C foil. This latter is introduced to minimize the contribution due to 8+ and 9+ beam charge states generated by the charge redistribution occurring in the primary target \cite{cavallaro2019} and elastically scattered within the spectrometer acceptance, which would produce a rate limitation for the FPD. To further reduce such contributions two aluminium screens were mounted at the entry of the FPD to intercept the unwanted ions. Such solution partially reduces the full momentum acceptance of the spectrometer and the number of hit silicon detectors. 
The experimental run was performed centering the spectrometer optical axis at $\theta_{lab}$ = 8$\degree$ corresponding to the angular acceptance $3\degree < \theta_{lab} < 14\degree$ in the laboratory frame. Thanks to the MAGNEX large (-14\%, +10\%) momentum acceptance, in addiction to the ($^{20}$Ne,$^{20}$O) DCE channel also the one-proton ($^{20}$Ne,$^{19}$F) and two-proton ($^{20}$Ne,$^{18}$O) transfers as well as the single charge exchange (SCE) ($^{20}$Ne,$^{20}$F) reaction were simultaneously detected with the same magnetic setting, corresponding to a magnetic rigidity value of 1.3894 Tm at the spectrometer optical axis.

\subsection{Particle identification purity}
\begin{figure}
\centering
\includegraphics[width=0.7\textwidth]{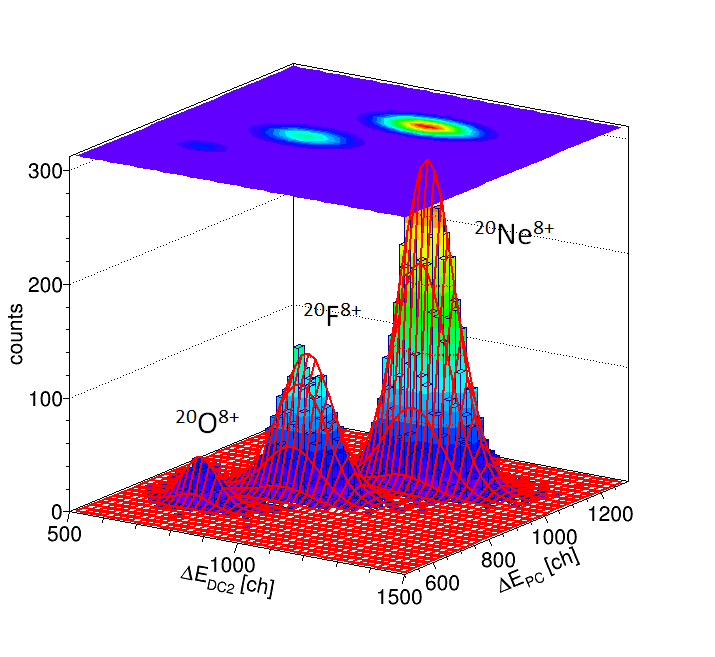} 
\centering \caption{(Colour online) Histogram of the $^{20}$Ne$^{8+}$, $^{20}$F$^{8+}$ and $^{20}$O$^{8+}$ events for one of the analyzed silicon detectors (No. 29) gated with the {$X_{foc}$ vs $E_{resid}$} topological selection of the $^{20}$O$^{8+}$ ions shown in Fig. \ref{Fig:1}(c). The superimposed global red line is the fit function described in the text.} 
\label{Fig:2}
\end{figure}

\begin{figure}[]
\centering
\includegraphics[width=0.49\textwidth]{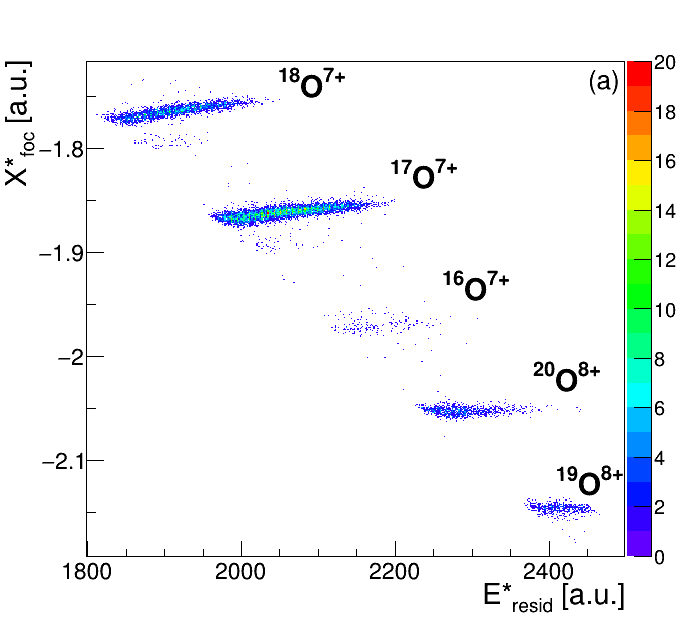} \includegraphics[width=0.49\textwidth]{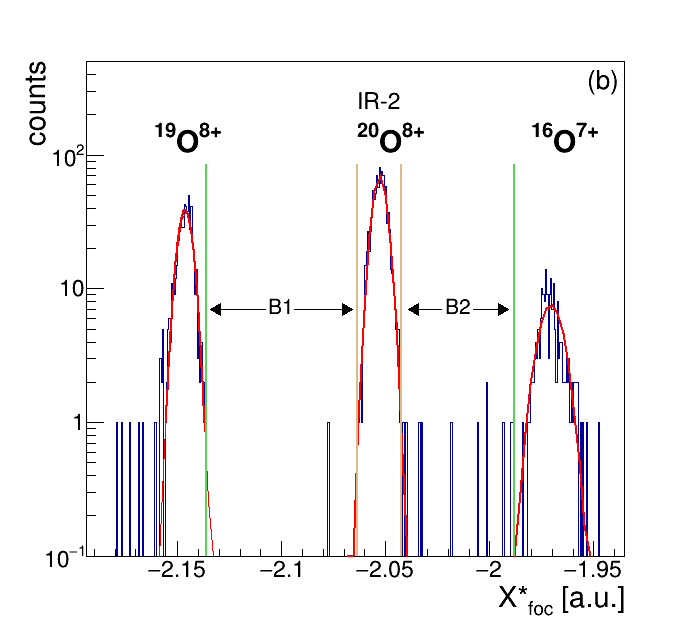} 
\centering \caption{(Colour online) Panel (a): \textit{$X_{foc}$ vs $E_{resid}$} correlation plot after an axis rotation for the events gated by the topological selection shown in Fig. \ref{Fig:1}(b) for the silicon detector No. 29. Panel (b): projection for the same events on the \textit{$X^*_{foc}$} axis. The fit functions described in the text are reported as well as the $^{20}$O$^{8+}$ ions identification region (IR-2).} 
\label{Fig:3}
\end{figure}

To estimate the purity, i.e. the discrimination capability of the described PID method in the selection of the DCE channel, the effect of the topological cut defined for the identification of the $^{20}$O$^{8+}$ ejectiles in the \textit{$X_{foc}$ vs $E_{resid}$} representation, as the one shown in Fig. \ref{Fig:1}(c), was first studied.
Since the $^{20}$O$^{8+}$ ions have very similar $\frac{\sqrt{m}}{q}$ ratio compared to $^{20}$F$^{8+}$ and $^{20}$Ne$^{8+}$ ones, thus sharing almost the same position in the \textit{$X_{foc}$ vs $E_{resid}$} correlation plot, such contour does not separate the three species (see Fig. \ref{Fig:1}(a)).   
Plotting the \textit{$\Delta E_{PC}$ vs $\Delta E_{DC2}$} histogram gated by the topological selection shown in Fig. \ref{Fig:1}(c), the three expected loci corresponding to the mentioned $^{20}$O$^{8+}$, $^{20}$F$^{8+}$ and $^{20}$Ne$^{8+}$ species are clearly visible as shown in Fig.~\ref{Fig:2}. Assuming bi-dimensional Gaussian models for the peaks, the histogram can be fitted by a three-Gaussian-sum function as shown in Fig. \ref{Fig:2}.
Then, the contribution of $^{20}$F$^{8+}$ and $^{20}$Ne$^{8+}$ events underneath the $^{20}$O$^{8+}$ peak is estimated by integrating the tails of their individual fits in the $^{20}$O$^{8+}$ identification region (IR-1). The latter corresponds to the elliptic domain defined by ($\pm$3$\sigma_x$, $\pm$3$\sigma_y$) from its centroid, consistent with the typical PID graphical selection width. The obtained misidentified events in the IR-1 region are listed in Tab.~\ref{tab:PIP} for those silicon detectors where the three analysed ejectiles are simultaneously detected due to their kinematic conditions. The errors were evaluated from the results of the fit, including also the correlation between the parameters. Upper limits within 68.27\% confidence level are reported when the relative error exceeds 100\% due to the limited statistics. 
\begin{table}[tp]
\caption{Impurities contribution from $^{20}$F$^{8+}$ (second column) and $^{20}$Ne$^{8+}$ (third column) in the $^{20}$O$^{8+}$ identification region (IR-1) for different silicon (Si) detectors (first column). The values are expressed in percentage with respect to the number of identified $^{20}$O$^{8+}$ ions. }\centering 
\begin{center}
\begin{tabular}{ccc}
\toprule
   Si No.    &  $^{20}$F$^{8+}$            & $^{20}$Ne$^{8+}$      \\ 
             &        (\%)                 &       (\%)            \\
\midrule
   20        &      $<$3.2$\times$10$^{-1}$  &       $<$5$\times$10$^{-4}$          \\
\midrule
   23        &      $<$2.7$\times$10$^{-1}$
                                             &   $<$3$\times$10$^{-8}$              \\
\midrule
   26        &      1.3$\pm$0.3$\times$10$^{-1}$      &       3$\pm$1$\times$10$^{-13}$   \\
\midrule
   29        &      1.2$\pm$0.1$\times$10$^{-1}$      &       8$\pm$4$\times$10$^{-15}$   \\
\bottomrule
\end{tabular}
\label{tab:PIP} 
\end{center}
\end{table}
These results represent the relative amount of misidentified $^{20}$O$^{8+}$ events from the \textit{Z} identification.  One readily notices that the contribution from $^{20}$Ne$^{8+}$ is considerably smaller than $^{20}$F$^{8+}$ one, so it will be neglected hereafter. \newline
About the \textit{A} and \textit{q} identification purity estimation, an analogous procedure was followed. The contour defined for the $^{20}$O$^{8+}$ in the \textit{$\Delta E_{PC}$ vs $\Delta E_{DC2}$} representation, as the one shown in Fig. \ref{Fig:1}(b), is adopted to explore the selected events in the \textit{$X_{foc}$ vs $E_{resid}$} correlation. In Fig.~\ref{Fig:3}(a) the thus gated events - already shown in Fig.~\ref{Fig:1}(c) for the same silicon detector - are presented after the following transformations: \textit{$E^*_{resid} = 0.9999 E_{resid}+0.0009 X_{foc}$}, \textit{$X^*_{foc} = -0.0009 E_{resid}+0.9999 X_{foc}$}. The latter are introduced to align horizontally the loci in order to project them on the new vertical axis (\textit{$X^*_{foc}$}). A zoomed view of the obtained mono-dimensional histogram is shown in Fig.~\ref{Fig:3}(b).
By repeating a similar procedure as that described to assess the purity in \textit{Z} - i.e. by fitting the structures visible in the mono-dimensional histogram as the one shown in Fig.~\ref{Fig:3}(b) with Gaussian functions - it emerges that the contribution to the $^{20}$O$^{8+}$ events from \textit{A} and \textit{q} identification is much smaller (average peak-to-peak distance of $\sim$20$\sigma$ of the corresponding fit functions) than that coming from the \textit{Z} identification. This is due to the better intrinsic resolution of both \textit{$X_{foc}$} and \textit{$E_{resid}$} measurements compared to \textit{$\Delta E_{DC2}$} and \textit{$\Delta E_{PC}$} ones \cite{cappuzzello2010,cappuzzello2011}.

\subsection{Particle identification background}              
Fig.~\ref{Fig:3}(b), however, highlights another important feature: few uncorrelated events are present between the well-separated peaks, not belonging clearly to any of them. In order to obtain a reliable background estimation of the PID technique, their contribution needs also to be taken into account.  
We focus the attention on those events present between the three peaks visible in Fig.~\ref{Fig:3}(b), corresponding to the ion of interest ($^{20}$O$^{8+}$) and its first two neighbors ($^{19}$O$^{8+}$ and $^{16}$O$^{7+}$) ones.
The spurious events located inside the B1 and B2 region of Fig.~\ref{Fig:3}(b) are considered as background. Such regions cover the intervals which extend more than 3$\sigma$ between two centroids. The background events do not appear to be distributed according to a clear pattern inside B1 and B2 region, thus a constant background linear density can be assumed. The amount of background events falling underneath the $^{20}$O$^{8+}$ selection region - defined as the interval spanning ${\pm}$3$\sigma$ around the $^{20}$O$^{8+}$ distribution centroid (IR-2, see Fig.~\ref{Fig:3}(b)) - is finally deduced from the estimated background linear density. The corresponding values, BG($^{20}$O$^{8+}$), expressed in percentage with respect to the number of identified $^{20}$O$^{8+}$ ions are listed in Tab.~\ref{tab:PIPs} and represent the background contributions for the different silicon detectors. The reported errors were estimated by Monte Carlo varying the limits of the B1 and B2 regions within the parameter errors resulting from the fits, repeating the density calculation and assuming as error one standard deviation of the obtained distribution. For silicon detector No. 20 the upper limit value within 68.27\% confidence level is reported.

\section{Estimation of the background equivalent cross section}
\begin{figure}[htb]
\centering
\includegraphics[width=0.7\textwidth]{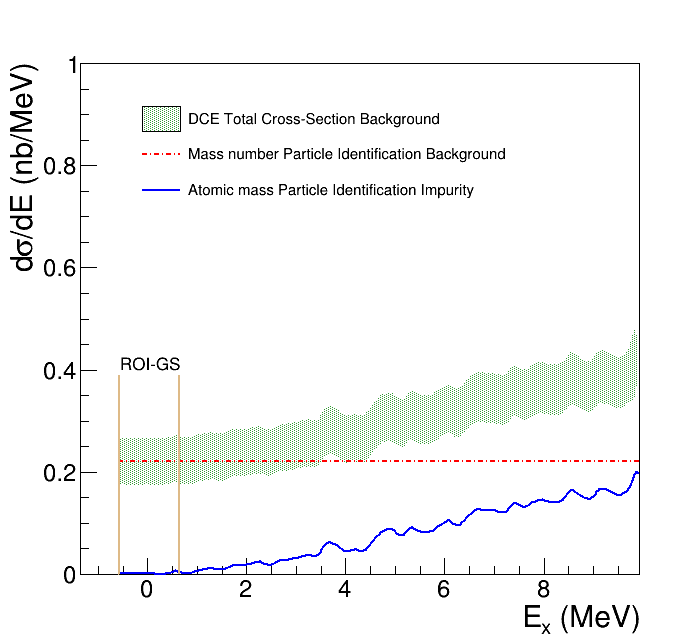} 
\centering \caption{(Colour online) Background equivalent cross section spectrum (green band) for the $^{116}$Cd($^{20}$Ne,$^{20}$O)$^{116}$Sn reaction at 15.3 AMeV as a function of the excitation energy. The PID background (red dotted line) and the $^{20}$F$^{8+}$ impurity (blue line) contributions as well as the ground-state region (ROI-GS) are also depicted. 1$\sigma$ error band is shown only for the total background cross section spectrum for clarity reasons.} 
\label{Fig:4}
\end{figure}

The results reported in the previous Section allow to estimate a background equivalent cross section which defines the MAGNEX spectrometer detection limit in the studied conditions. 
Regarding the PID impurity contribution in the \textit{$\Delta E_{PC}$ vs $\Delta E_{DC2}$} plot, the $^{20}$F$^{8+}$ contaminations listed in Tab. \ref{tab:PIP} are expected to be connected with the ($^{20}$Ne, $^{20}$F) SCE reaction cross section trend. In fact, the $^{20}$F$^{8+}$ ions represent the ejectiles produced in SCE reaction that was also simultaneously measured during the experimental run here described. Its preliminary analysis highlights a cross section growth at increasing excitation energy.  
The ratio between the estimated $^{20}$F$^{8+}$ events misidentified as $^{20}$O$^{8+}$ over the total $^{20}$F$^{8+}$ identified ones represents the estimated PID purity factor, R($^{20}$F$^{8+}$). The corresponding values, obtained for the different silicon detectors, are listed in Tab.~\ref{tab:PIPs}. They result quite similar, indicating that the trend of the $^{20}$F$^{8+}$ misidentified events in the DCE channel reflects that of the ($^{20}$Ne,$^{20}$F) SCE cross section spectrum.  
Thus, a constant factor given by the statistically-weighted average of the R($^{20}$F$^{8+}$) values - equal to 0.025$\pm$0.003\% - can be extracted. 
Scaling the ($^{20}$Ne,$^{20}$F$^{8+}$) SCE cross section spectrum by this value, the impurity contribution of $^{20}$F$^{8+}$ in the DCE one is estimated. The deduced contribution is shown in Fig.~\ref{Fig:4} as a function of the excitation energy ($E_x$) for the ($^{20}$Ne,$^{20}$O) DCE reaction.
\begin{table}[htp]
\centering \caption{PID purity factors of $^{20}$F$^{8+}$ ejectiles (R($^{20}$F$^{8+}$), second column) for selected silicon (Si) detectors (first column). The values are expressed in percentage with respect to the total number of identified $^{20}$F$^{8+}$ ions. PID background contribution to $^{20}$O$^{8+}$ events (BG($^{20}$O$^{8+}$), third column) in the IR-2. The values are expressed in percentage with respect to the number of identified $^{20}$O$^{8+}$ ions.} 
\begin{center}
\begin{tabular}{ccc}
\toprule
   Si No.    &  R($^{20}$F$^{8+}$)                &  BG($^{20}$O$^{8+}$)     \\
             &       (\%)                         &         (\%)             \\
\midrule
   20        &      $<$7.1$\times$10$^{-2}$       &     $<$1.9$\times$10$^{-1}$              \\
\midrule
   23        &      $<$6.7$\times$10$^{-2}$ 
   			                                      &     1.9$\pm$1.0$\times$10$^{-1}$         \\
\midrule
   26        &      2.8$\pm$0.6$\times$10$^{-2}$  &     2.8$\pm$0.2$\times$10$^{-1}$         \\
\midrule
   29        &      2.4$\pm$0.3$\times$10$^{-2}$  &     3.2$\pm$0.3$\times$10$^{-1}$         \\
\bottomrule
\end{tabular}
\label{tab:PIPs} 
\end{center}
\end{table} 
\newline
About the PID background contribution in the \textit{$X_{foc}$ vs $E_{resid}$} identification plot, it is likely related to not-exact residual energy measurement by the silicon detectors, due to possible phenomena like incomplete charge collection especially at their edges. However, since the corresponding values reported in Tab. \ref{tab:PIPs} for the different detectors are almost constant within the error bars, a unique factor given by the weighted average can be adopted ($<$BG($^{20}$O$^{8+}$)$>$ = 0.29$\pm$0.02\%) and a corresponding uniform cross section spectrum can be finally deduced. Such contribution is reported in Fig. \ref{Fig:4} and amounts to 0.22$\pm$0.04 nb/MeV. 
\newline
The equivalent cross section spectra of the two discussed contributions affecting the PID procedure, resulting from the particle identification background and impurity, are reported in Fig.~\ref{Fig:4} as a function of the system excitation energy for the ($^{20}$Ne,$^{20}$O) DCE reaction. Their sum, also shown in Fig.~\ref{Fig:4}, represents the overall PID background equivalent cross section spectrum of the $^{116}$Cd($^{20}$Ne, $^{20}$O)$^{116}$Sn reaction. The contribution from the spurious background events is dominant with respect to the impurity one which follows the SCE cross section increase. Such condition results favorable since the interest for the NUMEN project is mainly focused on the low excitation region of the spectrum, where isolated transitions to ground and first excited states should be visible. Of primary interest, in particular, is the ground-state to ground-state region (ROI-GS) which can be defined as $\pm$2$\sigma$ of the expected peak for the corresponding transition. Such interval determines the background equivalent cross section corresponding to the ground-state transition with a confidence level of 95.45\%. Wider confidence intervals would extent up to energy values where the corresponding cross section spectrum could be dominated by the transitions towards the first excited states. In the present case, assuming as expected resolution about 0.8 MeV at full width half maximum, the ROI-GS corresponds to $\pm$0.68 MeV. The estimated background value integrated in such region amounts to 0.30$\pm$0.03 nb. Such quantity finally determines the minimum cross section significantly measurable by the MAGNEX spectrometer, demonstrating that the DCE cross sections of interest for the NUMEN project are safely measurable. 
      
\section{Conclusions}
We have introduced a method to quantify the particle identification impurity and background of the MAGNEX magnetic spectrometer for the measurement of cross sections in heavy ion quasi-elastic reactions of interest for the NUMEN project. Combining their results it was deduced, for the first time, a procedure to estimate the expected background equivalent cross section spectrum. Such analysis is crucial because it provides quantitative indications of the minimum cross sections significantly measurable by the present MAGNEX spectrometer experimental set-up. The procedure was applied to the $^{116}$Cd($^{20}$Ne,$^{20}$O)$^{116}$Sn DCE reaction data but can be extended to any other system or reaction channel. In the presented case, the deduced cross section background value in the ground-state to ground-state region amounts to 0.30$\pm$0.03 nb. It should be noted that, similar results are expected considering the same reaction channels reported in the paper but with different heavy target systems and bombarding energies, as the ones adopted in the NUMEN experimental runs. In such conditions, both the FPD performances and the SCE reaction yields \cite{PhysRevC.98.044620} and, consequently, also the results achieved in this paper, are expected to be comparable. Moreover, the background equivalent cross section values here determined will guide the way toward the next experimental upgrade of the MAGNEX facility, foreseen by the next phase of the NUMEN project.

\section*{Acknowledgment}
This project has received funding from the European Research Council (ERC) under the European Union's Horizon 2020 research and innovation program (grant agreement No 714625). L.A. is grateful to the projects DGAPA-PAPIIT IN107820 and CONACYT LN-280769, LN-294537.
 
\bibliography{biblio}
\bibliographystyle{mprsty}
\end{document}